\documentstyle[epsf]{article}
\newcommand{\gsim}{\mbox{\raisebox{-.3em}{$\stackrel{>}{\sim}$}}}
\newcommand{\lsim}{\mbox{\raisebox{-.3em}{$\stackrel{<}{\sim}$}}}
\renewcommand{\cite}[1]{\ref{#1}}
\newcommand{\beq}{\begin{equation}}
\newcommand{\eeq}{\end{equation}}
\newcommand{\beqa}{\begin{eqnarray}}
\newcommand{\eeqa}{\end{eqnarray}}

\begin{document}
\baselineskip=0.6cm
\mbox{}\\[-3.5em]

\begin{center}
{\Large\bf Oklo Constraint on the Time-Variability of the Fine-Structure Constant}\footnote{Delivered at Astrophysics, Clocks and Fundamental Constants, 16-18 June 2003, Bad Honnef, Germany, to be published in {\sl Lecture Notes in Physics}, Springer}\\[.6em]
Yasunori Fujii\\[.6em]
{\small
\hspace*{-.4em}Advanced \hspace{-.2em}Research \hspace{-.2em}Institute \hspace{-.2em}for \hspace{-.2em}Science \hspace{-.2em}and \hspace{-.2em}Engineering, \hspace{-.2em}Waseda \hspace{-.2em}University,\hspace{-.2em} Tokyo, \hspace{-.2em}169-8555 \hspace{-.2em}Japan}
\\[.6em]
{\bf Abstract}
\end{center}
\baselineskip=.6cm
\begin{center}
\mbox{}\\[-2.2em]
\begin{minipage}{14.1cm}
The Oklo phenomenon, natural fission reactors which had taken place in Gabon about 2 billion years ago, provides one of the most stringent constraints on the possible time-variability of the fine-structure constant $\alpha$.  We first review briefly what it is and how reliable it is in constraining $\alpha$. We then compare the result with a more recent result on the nonzero change of $\alpha$ obtained from the observation of the QSO absorption lines.  We suggest a possible way to make these results consistent with each other in terms of the behavior of a scalar field which is expected to be responsible for the acceleration of the universe.
\end{minipage}
\end{center}
\mbox{}\\[-1.8em]

\section{What is the Oklo phenomenon?}

Oklo is the name of the place of a uranium mine in Gabon, West Africa, near the equator.  The  mining company would supply the uranium ore to the French government.  But in June of 1972, something unusual was noticed on the ore from Oklo; the abundance of $^{235}{\rm U}$ was somewhat below the world standard, 0.7202\%, well beyond the limit of permissible range.  This might have undermined the company's reputation about the quality of their uranium.  But finally after a few months of serious effort, French scientists came up with an unexpected, startling conclusion:

The deficit of $^{235}{\rm U}$ was a real effect of that a self-sustained fission reaction  took place {\em naturally} in Oklo about 1.8 Gys ago, during the period of Proterozoic, part of Precambrian.  In other words, natural reactors did exist well before 1942 when Enrico Fermi invented the artificial reactor for the first time in Chicago. This has been called the ``Oklo phenomenon," since then [\cite{naudet}].  The result of their work was published in many ways, including [\cite{IAEA}].\\[2em]
\hspace*{15em}
See the separate figure, 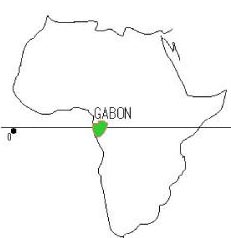. \\[.0em]

\begin{figure}[hbt]
\caption[]{Oklo in  Gabon, West Africa, near the equator
}
\label{gabon}
\end{figure}


There was a big press campaign, including the Le Monde article, for example, saying that Fermi was not an ``innovator," but was only an ``imitator" of Nature.

Even more surprising was that the occurrence of this ``natural reactors" had been predicted much earlier, 17 years earlier by a nuclear geochemist, Paul K. Kuroda in 1955 [\cite{kuroda}]. He discussed several conditions.  But the most important was  that the ratio of $^{235}{\rm U}$, currently 0.70\%, used to be much higher because of the different lifetimes of $^{235}{\rm U}$ and $^{238}{\rm U}$; $7.038 \times 10^8{\rm y}$ and $4.468\times 10^9{\rm y}$, respectively.  One can easily calculate  the ratio 1.8 Gys ago to be as high as 3.2\%.   We note that 3\% is a goal of most of today's enrichment facilities.  Another condition was the presence of water which served as a moderator.

So far 16 ``reactor zones" (RZ) have been discovered in the Oklo area.   In each of of them, extensive and detailed measurements have been made on the leftover fission products.

\section{How did Shlyakhter probe $\Delta\alpha$?}

Under this circumstance, in 1976, Alex Shlyakhter [\cite{nat},\cite{ATOMKI}] then in Leningrad proposed to look at $^{149}{\rm Sm}$, which is present naturally at the ratio 13.8\%, but should be depleted in the reactor zones because it had absorbed neutrons strongly in the reactors 2 Gys ago, according to the reaction
\begin{equation}
n+^{149}\hspace{-.2em}{\rm Sm} \rightarrow ^{150}\hspace{-.2em}{\rm Sm}+\gamma.
\label{ptb-1}
\end{equation}
One measures the abundance in Oklo reactor zones to estimate the cross section of this process, and compare the result with today's laboratory value.  In this way one can tell how much nuclear physics 2 Gys ago could have been different from what it is.

\begin{figure}[hbt]
\hspace{14.5em}
\epsfxsize=4.cm
\epsffile{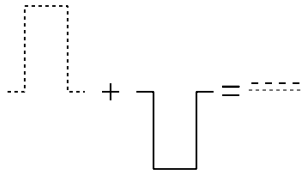}

\caption[]{Nearly complete cancellation between the repulsive Coulomb energy and the attractive nuclear energy, leaving a small leftover for the resonance energy}
\label{fig1}
\end{figure}

What is unique with this particular process (\ref{ptb-1}) is that it is dominated by a resonance that lies as low as $E_r = 97.3{\rm meV}$, while we know that a typical energy scale of nuclear physics is $\sim {\rm MeV}$.  Compared with this, the above value is very small, nearly 7 orders of magnitude too small.  This must be due to a nearly perfect cancellation between two effects; repulsive Coulomb force which is proportional to the fine-structure constant $\alpha$, and attractive nuclear force which depends on the strong-interaction coupling constant squared $\alpha_s$.  We are left with a very small leftover for the resonance energy, as illustrated Fig. \ref{fig1}.  Suppose we change one of the coupling constants, $\alpha$, say, only slightly.  Then the strength of the Coulomb energy will change also slightly, and so will $E_r$.    However, the {\em relative} change may not be so small, because the starting value $E_r$ was already small.  If this really happens, then the cross section may change rather significantly. This is a kind of amplification mechanism, which Shlyakhter exploited.

\begin{figure}[bht]
\hspace{12.em}
\epsfxsize=6.cm
\epsffile{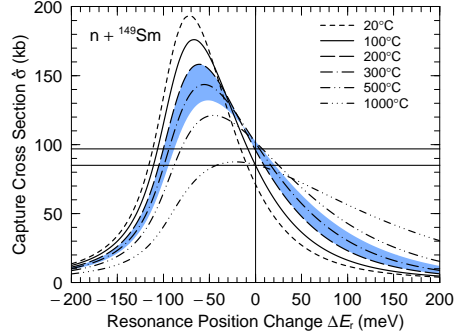}

\caption[]{The cross section $\hat{\sigma}_{149}$ for the process (\ref{ptb-1}) as a function of $\Delta E_r = E_r -97.3{\rm meV}$, assuming thermal equilibrium of the neutron flux.  The ranges of the observed cross section and the temperature are also shown, as given later by (\ref{ptb-7}) and (\ref{ptb-7-1}), respectively

}
\label{fig2}
\end{figure}

For the sake of illustration, we plot in Fig. \ref{fig2}  the cross section (based on the Breit-Wigner formula) as a function of $\Delta E_r$, the fictitious change of the resonance energy from today's value.  We also assume thermal equilibrium of the neutron flux; assuming a temperature corresponding to one of the curves shown.  We find a sharp peak obviously coming from the resonance. Suppose $E_r$ at 2 Gys ago were smaller by $10\:{\rm meV}$, a tiny amount.  Suppose also $T= 300^\circ{\rm C}$, for example.  Then we find the cross section bigger than today's value by a few \%, a significant change.

Incidentally, we are going to use the hat symbol attached to the cross section rather frequently. But this is only a technical  convention of normalization, which is particularly popular among the researchers of the Oklo phenomenon.  We do not worry too much at this moment.

Taking advantage of this strong dependence of the cross section $\sigma_{149}$ on $\Delta E_r$, Shlyakhter gave the upper bound
\begin{equation}
|\Delta E_r |\:\lsim\: 50\: {\rm meV}.
\label{ptb-2}
\end{equation}
It is not very much clear how he derived this result, particularly how much  the data uncertainties affected the conclusion. This is one of the points to be re-examined later.

He still went on to discuss how this change of $E_r$ corresponds first to the change of the strong-interaction coupling constant, $\alpha_s$.  He considered the resonance as a single-particle excitation in the potential, with its depth $V_0 \sim 50\:{\rm MeV}$,  which he assumed to be  proportional to $\alpha_s$.  If $\alpha_s$ changes, $V_0$ changes, and so does $E_r$.  Substituting from (\ref{ptb-2}), he obtained the result
\begin{equation}
\Biggl|\frac{\Delta\alpha_s}{\alpha_s}\Biggr|   = \Biggl| \frac{\Delta E_r}{V_0}  \Biggr| \:\lsim\: \frac{50\:{\rm meV}}{50\:{\rm MeV}}= 10^{-9},
\label{ptb-3}
\end{equation}
leading to the value $10^{-9}$.   Further dividing by $2\times 10^9{\rm y}$, he  arrived at
\begin{equation}
\Biggl|\frac{\dot{\alpha}_s}{\alpha_s}\Biggr| \:\lsim\: 5\times 10^{-19}{\rm y}^{-1}.
\label{ptb-4}
\end{equation}

As for the electromagnetic interaction, he apparently replaced $\alpha_s$ by $\alpha$, resulting in dividing these by $ \alpha/\alpha_s \sim 1/20$, giving
\begin{equation}
\Biggl|\frac{\Delta\alpha}{\alpha}\Biggr| \:\lsim\: 2\times 10^{-8},\quad
\Biggl|\frac{\dot{\alpha}}{\alpha}\Biggr| \:\lsim\: 1\times 10^{-17}{\rm y}^{-1},
\label{ptb-5}
\end{equation}
This last value has been several orders of magnitude more stringent than any other estimates, a kind of ``champion result"  for many years. However, one may raise a question against the argument from $\alpha_s$ to $\alpha$, and the suspicion may go further back to the derivation (\ref{ptb-3}).

For better understanding, we re-examined the whole analysis [\cite{yfetal}] by forming a team, which includes theorists of nuclear physics, reactor scientists and geologists.  Among them, Hiroshi Hidaka is an expert nuclear chemist who has  been specialized to the Oklo phenomenon.

\section{How good is it?}

We soon realized that a major error source of the data comes from the ``post-reactor contamination," implying that certain amount of $^{149}{\rm Sm}$ present in the outside environment flowed into the reactor core having occurred after the end of the reactor activity that is believed to have lasted several $10^5$ years.  This amount has nothing to do with what happened inside the reactors, so is a contamination from our purpose.  This inflow was in fact the gradual mixture between inside and outside prompted by repeated successions of dissolution and precipitation of Sm, caused essentially by weathering, namely being exposed to the air.  To minimize this  embarrassing effect, we looked for samples in the reactor zones 10--16 discovered later than 1984, deep underground, as shown  in Fig. \ref{fig3}.  Finally we decided to collect five samples taken from RZs 10 and 13 below the surface, with enough care of geologists expertise. \\[2em]
\hspace*{15em}
See the separate figure, 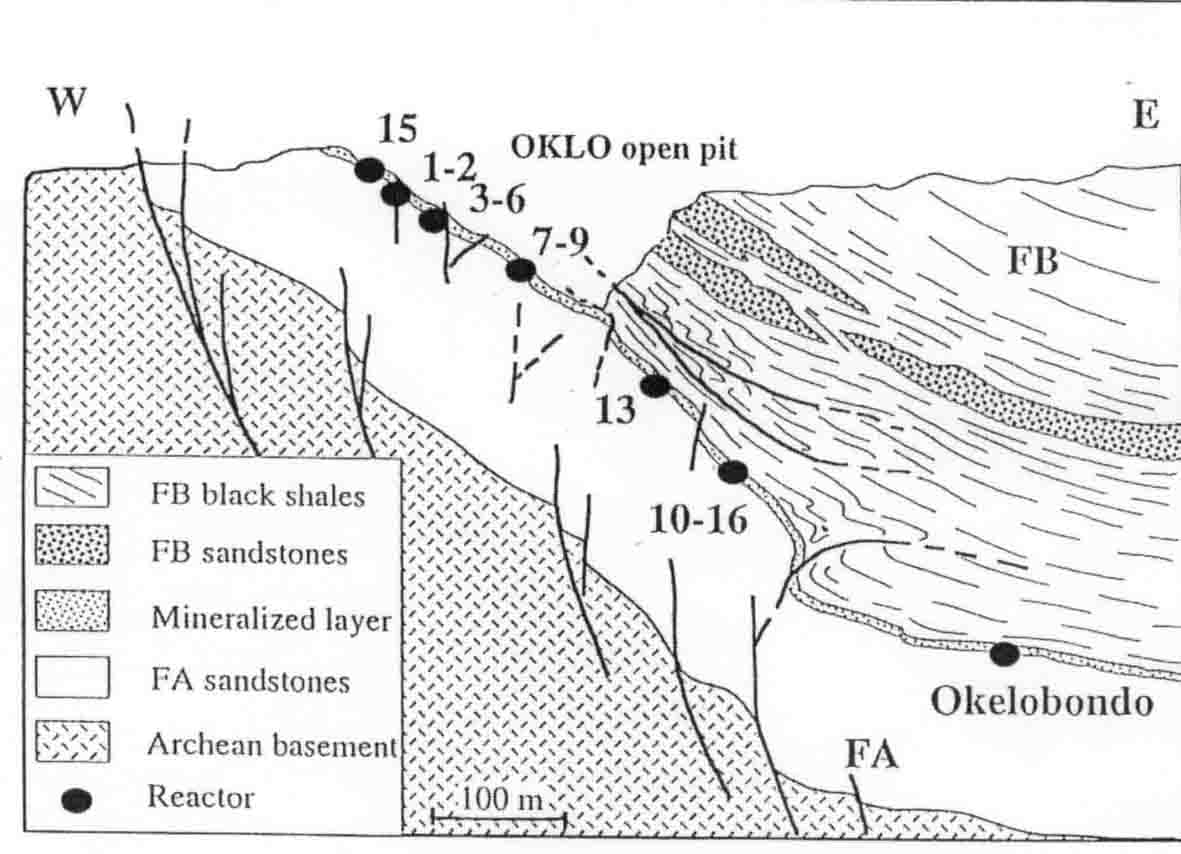. \\[.0em]

\begin{figure}[hbt]
\caption[]{Geological vertical cross section of the Oklo area
}
\label{fig3}
\end{figure}

The measured isotopic ratios related to $^{149}{\rm Sm}$ for five  samples are shown in Table 1.  We see how small the abundances of $^{149}{\rm Sm}$ are  compared with the natural abundance 13.8\%.  We did not show the errors, but they are simply small. $t_1$ is the time of the end of the reactor activity that started at $t=0$.  Also in the first line, we entered what is called ``fluence," denoted by $\hat{\phi} t_1$, but actually a time-integrated neutron flux $\hat{\phi}$ during the whole duration of reactor activity.

\begin{table}[tbh]
\caption{Measured isotopic ratios related to $^{149}{\rm Sm}$ obtained from five samples.  The fluence and cross section are also listed
}
\begin{center}
\begin{tabular}{ c c  c c r r r r r c} \hline
& & & & & & & & &\\[-0.07in]
& & & \multicolumn{7}{c}{Sample} \\[0.05in]
&  & & & SF84-1469 &SF84-1480 & SF84-1485 & SF84-1492 & SD.37& \\
& & & & & & &&& \\[-0.07in]
\hline
& & & & & & & &&\\[-0.07in]
&$\hat{\phi}t_1$ (1/kb)&  & & 0.525 & 0.798 & 0.622 & 0.564 & 0.780&  \\
&$N_{144}(t_1)$ (\%)  &  & &  0.1052  & 0.2401 & 0.2073 & 0.1619 & 0.06909 &\\
&$N_{147}(t_1)$ (\%)  &  & & 55.34 & 53.23 & 54.03 & 54.81 & 52.74 &\\
&$N_{148}(t_1)$ (\%)  &  & & 2.796 & 3.468 & 3.079 & 2.890 & 4.694& \\
&$N_{149}(t_1)$ (\%)  &  & & 0.5544 & 0.2821 & 0.4466 & 0.4296 & 0.3088 &\\
&$N_{235}(t_1)/N_{238}(t_1)$ &  &&0.03181 & 0.02665 & 0.02971 & 0.03047 & 0.02435& \\[0.08in]
\hline
& & & & & & & &&\\[-0.07in]
&$\hat{\sigma}_{\rm 149}$ (kb) & & & 85.6 & 96.5 & 83.8 & 99.0 & 89.5 &\\[0.08in]\hline
\end{tabular}
\end{center}
\end{table}

We then solved the evolution equations 
\begin{eqnarray}
dN_{147}(t)/dt&=&-\hat{\sigma}_{147}\hat{\phi}N_{147}(t) +
N^0_{235}\exp (-\hat{\sigma}_{\rm a}\hat{\phi}t) \hat{\sigma}_{\rm f235}\hat{\phi}Y_{147}, \nonumber\\
dN_{148}(t)/dt&=&\hat{\sigma}_{147}\hat{\phi}N_{147}(t), \nonumber\\
dN_{149}(t)/dt&=&-\hat{\sigma}_{149}\hat{\phi}N_{149}(t) +
N^0_{235}\exp (-\hat{\sigma}_{\rm a}\hat{\phi}t) \hat{\sigma}_{\rm
f235}\hat{\phi}Y_{149}. \nonumber
\end{eqnarray}
for the related isotopes to calculate the cross section $\hat{\sigma}_{149}$ for the process (\ref{ptb-1}).  The result is summarized,  
\begin{equation}
\hat{\sigma}_{149} = (91\pm 6)\:{\rm kb},
\label{ptb-7}
\end{equation}
corresponding to the narrow horizontal band shown in Fig. \ref{fig2}.

We also made an estimate of the temperature, by the traditional way supplemented by a latest technique, giving 
\begin{equation}
T=(200-400)^\circ{\rm C},
\label{ptb-7-1}
\end{equation}
corresponding to the shaded area in Fig. \ref{fig2}.
We find two intersections, and the corresponding two separated ranges of $\Delta E_r$.  
\begin{equation}
\Delta E_r =
\left\{
\begin{array}{ll}
\hspace{.5em}(9\pm 11)\:{\rm meV}, &\quad\mbox{right-branch, Null}\\[.6em]
(-97\pm 8)\:{\rm meV}, &\quad\mbox{left-branch, Non-Null}
\end{array}
\right.
\label{ptb-10}
\end{equation}

The right-branch range covers zero, so that a null result in the usual sense, while the other implies that $E_r$ was different from today's value by more than 10 standard deviations.  Does this really imply an evidence of the difference in 2 billion years ago?  We tried to see if the non-null result can be eliminated by looking at other isotopes like $^{155,157}{\rm Gd}$, but so far no final conclusion yet.

At this point we compare our result with those due to Damour and Dyson [\cite{dd}] (DD), who used the samples obtained  mainly from near the surface, giving   the cross section:
\begin{equation}
\hat{\sigma}_{149} = (75\pm 9){\rm kb},
\label{ptb-8}
\end{equation}
somewhat smaller than our result (\ref{ptb-7}).  This seems  consistent with our suspicion that their data suffered from contamination.  Also, they did not come to separating the two ranges. They could have done it, though the ``right-branch" range would failed to cover zero even at the level of 2 standard deviations.  Instead, they gave only a combined range, $-120\:{\rm meV}\:\lsim\:\Delta E_r\:\lsim\: 90\:{\rm meV}$, which more than covers our two ranges.

We admit that we are still short of determining which range  is correct. Then one might say that we should also be satisfied by the combined range.  But we still insist that it is a progress to have established a disallowed range in between.

Now we move on to discuss how $\Delta E_r$ is translated into $\Delta\alpha$ following DD.  First they ignored the contribution from the strong interaction entirely, focusing on the first ``term" coming from the Coulomb contribution in Fig. \ref{fig1}.

Consider the energy ${\cal M}_c$,  given by the difference of the Coulomb energies between the states with 150 and 149.  They paid special attention to the fact that the resonance in the $^{150}{\rm Sm}$ is excited.  But we simplify the analysis, at this moment, by appealing to the semi-empirical mass formula due to Weizs\"{a}cker, finding ${\cal M}_c \approx -1.1 {\rm MeV}$.  In this calculation, one has to allow an error perhaps within the factor 2 or 1/2.   Notice also that the above result is negative, apparently in contradiction to the illustration in Fig. \ref{fig1}.  Obviously we dealt basically with a repulsive force, but we calculated the difference, which turns out to be negative.  Nothing is wrong, but we would better put the two terms upside-down on the left-hand side, but  keeping the right-hand side still positive.  We may also assume that ${\cal M}_c \propto \alpha$. We then obtain $\Delta E_r = \Delta {\cal M}_c =(\Delta\alpha/\alpha){\cal M}_c$, thus giving
\[
\frac{\Delta\alpha}{\alpha} =\frac{\Delta E_r}{{\cal M}_c}=\left\{
\begin{array}{ll}
(-0.8 \pm 1.0)\times 10^{-8}, &\mbox{null, upper bound} \nonumber\\
(0.88 \pm 0.07)\times 10^{-7},&\mbox{non-null} 
\end{array}
\right.
\]
Divide by $-2\times 10^9{\rm y}$ to get
\[
\frac{\dot{\alpha}}{\alpha}=\left\{
\begin{array}{ll}
(0.4\pm 0.5)\times 10^{-17}{\rm y}^{-1}, &\mbox{null, upper bound}\\
(-0.44\pm 0.04) \times 10^{-16}{\rm y}^{-1},&\mbox{non-null}
\end{array}
\right.
\]
This upper bound happens to agree quite well with Shlyakhter's result $1\times 10^{-17}{\rm y}^{-1}$.  The agreement to this extent seems, however, rather accidental, because, among other things, it is unlikely that the data as good as ours was available in 1976.

We emphasize here that the simple estimate due to DD, as described here, might be called ``Coulomb-only estimate," which serves as a basis for more general analyses.

In fact what really happens might be a combined result of both interactions, and one wishes if one could include the strong interaction as well. But then  everything is going to be complicated, for example, like the QCD analysis in [\cite{olive}].   But there are something independent of such complications as long as we appreciate the condition that $\Delta E_r$ is much smaller than either of the mass scales, ${\cal M}_c$ and its strong-interaction counterpart ${\cal M}_s$.  First we find from Fig. \ref{fig1} that the mass scale of ${\cal M}_s$ is nearly equal in its size to that of ${\cal M}_c$,  obviously much smaller than Shlyakhter's ``$50\:{\rm MeV}$."  It then also follows that $\Delta\alpha_s/\alpha_s$ should be nearly of the same size as $\Delta\alpha/\alpha$.  Of course there are some differences from the Coulomb case; ${\cal M}_s$ may not be simply proportional to $\alpha_s$.  This may result in a revision of a factor, but certainly not of an order of magnitude.  Then we go through a bit of analysis to conclude finally that it is unlikely that, by the strong interaction, $\Delta\alpha/\alpha$ deviates from the Coulomb-only estimate by more than an order of magnitude, no matter how complicated the exact analyses might be.  It can be  smaller.  See Appendix A for more details.

Then, as always, there is a possibility of an exception, no matter how remote.
This allows, in principle, that both of $\Delta\alpha/\alpha$ and $\Delta\alpha_s/\alpha_s$ are quite large, in fact without limit, but cancel each other leaving a small value of $\Delta E_r$.  At this moment, however, we assume that no such fine-tuning nor coincidence occurs in the real world.

Let us summarize what the situation is with the question of ``uncertainties."  
In the theoretical aspect as a whole, we say again that the Coulomb-only estimate of the relative change of $\alpha$ is correct likely within an order of magnitude, for whatever the complication of the effect of strong interaction. We only add a few related remarks.
\begin{itemize}
\item Only $\Delta E_r$ and the related cross section $\sigma_{149}$ are sensitive to $\Delta\alpha$.  No other quantities are.
\item Our formulation is such that, to a good approximation in practice, the neutron flux can be any function of time. For example, the reactor activity can be even ``sporadic."  The only thing that counts is the fluence, no matter how long it took.
\item Estimating fluence is complicated but is a standard estimate and is reliable.
\item In principle we may not rule out one of the higher resonances to come down near the threshold, giving much larger value of $\Delta E_r$.  We have a reason, however, to believe this to be highly unlikely.  See Appendix B. 
\end{itemize}

In the observational aspect, we repeat our previous statement; post-reactor contamination is the largest error source, with a few more comments;  
\begin{itemize}
\item Nothing is serious for $^{149}{\rm Sm}$ in our samples.  (A few \% contamination seems even better because then the range of $\Delta E_r$ for the null result covers zero more in the middle.)
\item This is not the case for $^{155,157}{\rm Gd}$ which enjoy even lower resonance energies.  The absorption cross sections are also larger.  But the effects are too strong to the extent that the residual abundances are too little, so are too sensitive to contamination, even with our samples with minimized effect of weathering.  Shlyakhter was clever, when he chose $^{149}{\rm Sm}$.  This is also precisely why we reached short of complete elimination of the non-null result, as mentioned before.  
\end{itemize}


\section{How can it be consistent with the QSO result?}

\begin{figure}[tbh]
\hspace{9em}
\epsfxsize=7.cm
\epsffile{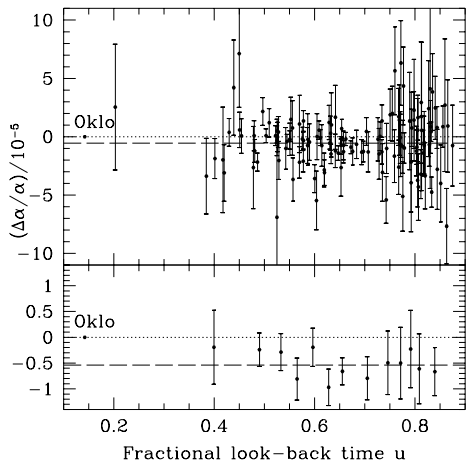}

\caption[]{QSO result from the 128 data points is shown in the upper panel, while the 13 binned data might provide an easier view in the lower panel [\cite{ww2}].  The long-dashed line is for the weighted mean $-0.54$.  The Oklo time $u_{\rm oklo}=0.142$ is also shown}
\label{fig4}
\end{figure}

According to V. Flambaum and M. Murphy at this meeting, the latest version of their result on the time-variation of $\alpha$ from spectroscopy of QSO absorption systems is [\cite{ww2}] (see also their contribution to the proceedings [\cite{ww3}]):
\[
\frac{\Delta\alpha}{\alpha}= (-0.54\pm 0.12) \times 10^{-5}.
\]
We show in Fig. \ref{fig4}, taken from Fig. 8 of [\cite{ww2}],  our own plot as a function of the fractional look-back time $u$  defined by $u=(t_0 -t)/t_0$, with $t_0$ the present age of the universe.  Their weighted mean can be viewed as a fit by a horizontal straight line at $-0.54$, as also shown in Fig. \ref{fig4}.  We will call this a ``1-parameter fit" for the later convenience. Notice that $\chi^2_{\rm red}= 1.06$. 

We now include the data from the Oklo phenomenon, at $u_{\rm oklo}=0.142$, barely outside the QSO range, as also shown in Fig. \ref{fig4}.  In this sense they are different things.  However, we should  put  error-bars, which are invisibly small in this plot; less than $10^{-4}$ and $10^{-2}-10^{-3}$ in the horizontal and vertical directions, respectively.  Remember that $\Delta\alpha/\alpha$ from the Oklo constraint is $10^{-7}-10^{-8}$. If we extend the straight line naively down to the Oklo time, it will miss the point off $(10^2-10^3)$ standard deviations, resulting in an enormous value of $\chi^2$, too large to be acceptable.  One wants to bend the line to pass the point almost exactly, but one needs a physical reason. What is that? This is the issue. Already there have been  several attempts [\cite{BM1}--\cite{gardner}]. But we stick to our own idea that this issue has something to do with the accelerating universe, another big issue in today's cosmology.

Now probably everyone knows that our universe is accelerating [\cite{randp}].  This behavior is best described in terms of a positive cosmological constant, whose size is given usually by the parameter $\Omega_\Lambda=\Lambda/ \rho_{\rm cr}\sim 0.7$, where the critical density $\rho_{\rm cr}$ is given by $\sim t^{-2}_0$.  The coefficient here is of the order one if we use the reduced Planckian unit system in which $c=\hbar= M_{\rm P}(= \sqrt{8\pi G/(c\hbar)})=1$.  In this unit, the present age $t_0 \sim 1.4\times 10^{10}{\rm y}$ is about $10^{60}$.  So we find $\rho_{\rm cr}\sim \Lambda \sim 10^{-120}$.

Today's cosmological constant problem has two faces or questions: Why is it so small? Why is it still nonzero?  The first question can be replied by the ``scenario of a decaying cosmological constant;" $\Lambda$ is not a true constant but decays like $\sim t^{-2}$ [\cite{yfdcl},\cite{dolgov}].  This simple idea can be implemented by the ``scalar-tensor theory."  We expect that a scalar field  plays an important role.  This scalar field may have its origin in string theory in which a graviton has a spinless companion called  the ``dilaton."

The second question seems to require a deviation from the simplest version of the scalar-tensor theory.  As one of the possible ways we call for another scalar field, called $\chi$, in addition to the dilaton $\sigma$.  These two fields comprise what is called the ``dark energy," and their energy density $\rho_s = \rho(\sigma, \chi)$ is interpreted as an effective cosmological constant $\Lambda_{\rm eff}$.  There are many details involved, though we are not going into any details.  Readers are advised to refer to our recent book [\cite{cup}]. 
\begin{figure}[tbh]
\hspace{10em}
\epsfxsize=7.cm
\epsffile{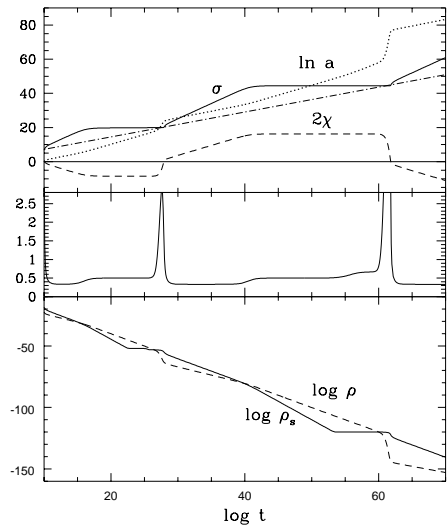}

\caption[]{An example of our cosmological solutions in the Friedmann universe wit $k=0$.  The scale factor $a$, the scalar fields $\sigma$ and $\chi$, the energy densities $\rho$ and $\rho_s$ of the ordinary matter and the scalar fields, respectively, are plotted against $\log t$, in the reduced Planckian unit system.  The present time is around $\log t\approx 60$.  The middle panel shows an effective exponent of the scale factor, $\ln a/\ln t$
}
\label{fig5}
\end{figure}

Skipping all the details, we show in Fig. \ref{fig5} an example of our solutions in the Friedmann universe with flat 3-space.  The horizontal axis is $\log t$.  In the Planckian unit system, the present time is somewhere around 60.  In the lower panel, $\rho$ is the usual matter energy density, which falls off roughly as $t^{-2}$.  The energy density of the scalar fields $\rho_s$ is the effective cosmological constant, $\Lambda_{\rm eff}$, also falling off like $t^{-2}$ as an overall behavior, thus respecting the scenario of a decaying cosmological constant.\footnote{As emphasized in [\cite{cup}], the gravitational ``constant" $G$ in the {\em physical} conformal frame identified (nearly) with the Einstein frame in the this model is (nearly) time-independent, instead of decaying with time as in [\cite{dolgov}] presented in the Jordan frame}  But the plot also shows occasional deviations, notably the plateau behaviors.  Obviously each plateau mimics a cosmological ``constant."  Furthermore, it comes to a crossing with the ordinary matter energy density.  One of them is expected to occur around the present epoch.  Nearly in coincidence with this crossing, we find a ``mini-inflation" of the scale factor; a bit of sharp increase in $\ln a$ shown in the upper panel.  This nicely fits the observed acceleration of the universe.

\begin{figure}[tbh]
\hspace{10em}
\epsfxsize=6.cm
\epsffile{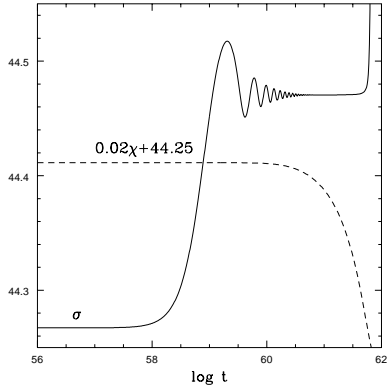}

\caption[]{A close-up view of $\sigma$ and $\chi$ in the upper panel of Fig. \ref{fig5} around the present epoch.  The magnification rate in the vertical direction is 330
}
\label{fig6}
\end{figure}

Also shown in the upper panel are the sudden changes of the scalar fields $\sigma$ and $\chi$, again in coincidence with the crossing between $\rho$ and $\rho_s$.  However, the most  interesting is to take a close-up view of what appears to be a simple and small jump of $\sigma$.  With the magnification rate as large as 330 in the vertical direction, we find a surprising behavior shown in Fig. \ref{fig6}, something like a damped oscillation.  

We know what the underlying mechanism is at the deeper level. But we still say that this damped-oscillation-like behavior is in fact ``the heart and soul" of the entire dynamics that eventually brings about the acceleration as we see it.  On the other hand, the acceleration itself does not care how invisibly tiny oscillation is taking place behind the scene.  There are many variations in the way  of oscillation.  In this sense, we have a degeneracy, which the cosmological acceleration does not resolve.  However, this invisibly small oscillation may show up through the time-variation of $\alpha$.

This is an expectation based on a general view that changing $\alpha$ if any is due to the changing scalar field, expressed symbolically as
\begin{equation}
\frac{\Delta\alpha}{\alpha}\propto \Delta\sigma .
\label{ptb-11}
\end{equation}
String theory suggests this dependence for the gauge coupling constant.  We ourselves derived a relation of this type, based on QED, featuring a quantum-anomaly type of calculation.  But we do not want to be too specific on these theoretical details, nor to depend heavily on the choice of the solutions, like the one in Fig. \ref{fig6}. This is particularly crucial because we have many different solutions for a given cosmological behavior we want to fit.  Rather, we are going to follow a phenomenological approach which we describe briefly, leaving more details to [\cite{yfplb}].

Let $y$ denote $\Delta\alpha/\alpha$ in units of $10^{-5}$.  Then we assume a dependence on the fractional look-back time $u$ in the way of a damped oscillation
\begin{equation}
y= ae^{bu}\sin \left( 2\pi \frac{u-u_{\rm oklo}}{T} \right),
\label{ptb-12}
\end{equation}
where the parameters are going to be determined to fit the QSO data as well as the Oklo constraint.

We first choose $u_{\rm oklo}=0.142$ corresponding to the Oklo time of $1.95\:{\rm Gys}$ ago.  The Oklo constraint, to be $10^{-2}-10^{-3}$ in terms of $y$, is approximately zero in this scale.  The remaining parameters $a,b,T$ are determined by minimizing $\chi^2$ for the QSO data.  In this sense we call this a 3-parameter fit. We do not include the Oklo data in computing $\chi^2$, because we consider the Oklo has been already fitted approximately by choosing a zero of the function as above.
\begin{figure}[htb]
\hspace{9.em}
\epsfxsize=7.cm
\epsffile{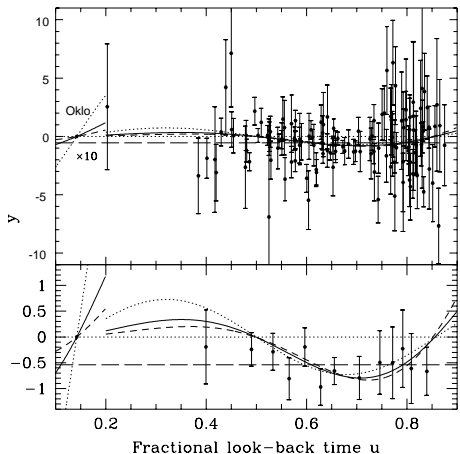}
\caption[]{The solid curves are for the 3-parameter fit with the least $\chi^2$ $(a= 0.151, b=2.4, T= 0.714, \chi^2_{\rm red} =1.09)$, to be compared with the 1-parameter fit, represented by a horizontal straight (long-dashed) line at $-0.54$ in Fig. \ref{fig4}.  Note the 10 times magnification for $u<0.2$.   The dotted and dashed curves are for $b=0.0$ and $b=4.0$, respectively
}
\label{fig7}
\end{figure}

We limit ourselves to a region of $a,b,T$ in a manner roughly consistent with the theoretical model of the accelerating universe.  In this range we searched for  local minima of $\chi^2$.  Among several of them we find the least minimum which is given by $b=2.4, a= 0.151, T= 0.714$ resulting in $\chi^2_{\rm red}= 1.09$. This $\chi^2_{\rm red}$ is similar to $\chi^2_{\rm red}= 1.06$ obtained for their 1-parameter fit. In this sense our 3-parameter fit is nearly as good as the fit in [\cite{ww2}].  The solid curve in Fig. \ref{fig7} shows the actual plot, probably better shown in the binned plot in the lower panel.  We magnified the curves below $u=0.2$ by 10 times.

One might ask us why we are satisfied with $\chi^2_{\rm red}=1.09$ which is not smaller than 1.06 for the 1-parameter fit, in spite of the fact that we have more degrees of freedom. We answer the question by pointing out the following:
\begin{itemize}
\item Our $\chi^2_{\rm red}=1.09$ is for the whole data including the Oklo, because, as we noted, the Oklo constraint has been already ``included" in a sense.  In comparison, however, the 1-parameter fit gives an unacceptably large $\chi^2$ when we include the Oklo, which was the starting point of the whole discussion.
\item Our 3-parameter fit was motivated originally by a theoretical ``prejudice."  There was no guarantee that it fits the reality.  We are relieved to find that  our prejudice somehow survived a realistic test.
\end{itemize}

\begin{figure}[tbh]
\vspace{-7em}
\hspace{9em}
\epsfxsize=7.cm
\epsffile{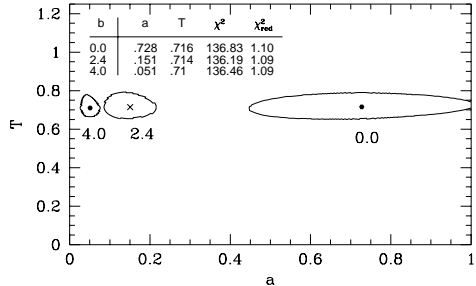}

\vspace{-1.6em}
\caption[]{The 3-dimensional 68\% confidence region is illustrated in terms of three cross sections for $b=0.0, 2.4, 4.0$, as marked beside each contour, shown in 2-dimensional $a-T$ space 
}
\label{fig8}
\end{figure}

We now discuss what the confidence region for $68\%$ is like for this 3-parameter fit.  We imagine a confidence volume in 3-dimensional space of $b, a, T$.  Figure \ref{fig8} shows, however, 2-dimensional cross sections for three different values of $b$.  The contour labeled by 2.4 in this figure shows the one for $b=2.4$.  We also show other 2 cross sections corresponding to $b=0.0$ and $b=4.0$, respectively.  They give only slightly larger $\chi^2$ than that for $b=2.4$.  In this way one imagines what the 3-dimensional volume looks like.  The curves for these $b$ are also plotted by the dotted and dashed curves, respectively, in Fig. \ref{fig7}.  They are different from each other only in the lower-$u$ region, $u\:\lsim\: 0.5$.

We further add that we obtained several other solutions with other values of $b, a, T$ which give local minima of $\chi^2$, as we indicated before.  As it turned out, however, they tend to give $\chi^2_{\rm red}\gsim 1.2$.  This is a number which is nearly comparable with $\chi^2_{\rm red}=1.24$, which we would obtain by fitting the QSO data by a horizontal straight line $y=0$, namely the $u$-axis itself.  We may have a good reason to exclude these fits.

We may compare the result shown in Fig. \ref{fig7} with the cosmological solution which we started from, as we showed in Fig. \ref{fig6}.  For the latter we may estimate the parameters approximately, which will be shown in Table 2, together with the corresponding ones for the former.

\begin{table}
\caption{Comparison between the damped-oscillator-like fit with an example of the cosmological solution}
\begin{center}
\renewcommand{\arraystretch}{1.4}
\setlength\tabcolsep{5pt}
\begin{tabular}{cccc}
& $a$ & $b$ & $T$ \\
\hline
Cosmological solution & \raisebox{-.5em}{$\approx 2.4$} & \raisebox{-.5em}{$\approx 2.5$} & \raisebox{-.5em}{$\approx 0.22$} \\
\raisebox{.5em}{in Fig. \ref{fig6}} & & & \\
Fit in Fig. \ref{fig7}& $0.15\pm 0.05$ & $2.4\pm 2$ & $0.71\pm 0.06$ \\
\hline
\end{tabular}
\end{center}
\label{Tab1a}
\end{table}

\begin{figure}[tbh]
\hspace{9em}
\epsfxsize=6.cm
\epsffile{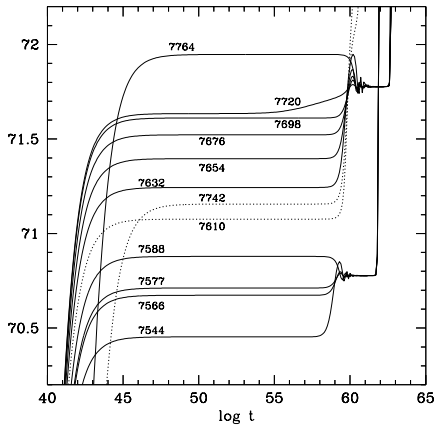}

\caption[]{Another magnified view of the behavior of $\sigma$.  The previous Fig. \ref{fig6} is only a small part of the present figure, corresponding to a curve with the attached number $7544$.    These numbers show the last 4 digits in the initial value at $t=10^{10}$, as explained in detail in [\cite{cup}].  The nucleosynthesis and CMB times are $\approx 45$ and $\approx 55$, respectively.  The vertical scale may be accepted arbitrary at this moment
}
\label{fig9}
\end{figure}

Agreement in the values of $b$ is obvious.  We have to have more theoretical details in order for the comparison of $a$ to make sense, though we may reasonably find a consistency.  On the other hand, there is a discrepancy between the values of $T$.  The available QSO data shows a rather flat distribution of $\Delta\alpha/\alpha$, which favors a ``larger" $T$.  In this connection we point out, however, that we have chosen the solution in Fig. \ref{fig6} rather arbitrarily.  In fact Fig. \ref{fig9} indicates that it happened that we have come across a relatively small $T$.  We only conclude that we have to look for other solutions of the cosmological equations which still fit the way of the cosmological acceleration.

Finally, we add the following comments for further improvements of the fit.  
\begin{itemize}
\item We show Fig. \ref{fig9}, taken from Chapter 5 of [\cite{cup}], which includes Fig. \ref{fig6} as a small portion for a special choice of the initial value. This figure also demonstrates that the detailed behavior of the small oscillation depends heavily on the early history, depending sensitively on the behavior of the scalar fields at the initial times, particularly on those around the time of nucleosynthesis. In this sense determining $\Delta\alpha/\alpha$ in this epoch as well as in the CMB era is crucially important in this approach.  
\item In our approach in terms of (\ref{ptb-12}) we assumed that the present time corresponds to the limiting, still transient, behavior toward the common flat value immediately prior to the big and sudden jump of $\sigma$.  It seems better to consider that we are in the middle of the oscillation behavior, in general.  Taking this possibility into account will make it more likely to satisfy the natural condition $y(0)=0$, which should be true by definition.  This type of the fit will be discussed in Appendix C.
\item Our analysis is based on the simplest assumption on the $\sigma-\chi$ interaction, as given by (5.58) of [\cite{cup}].  This might be modified to improve the fit.

\end{itemize}

As the very last comment, we hope if natural reactors will be discovered somewhere else, thus providing us with additional constraints, hopefully at different times. \\[1em]

\noindent
{\large\bf Appendix}\\[-.5em]
\appendix
\renewcommand{\thesection}{\Alph{section}}
\renewcommand{\theequation}{\Alph{section}.\arabic{equation}}
\setcounter{section}{0}
\setcounter{equation}{0}
\section{Bound on $\Delta\alpha/\alpha$ from the Coulomb-only estimate}

The situation described in Fig. \ref{fig1} may be given the expression
\begin{equation}
E_r ={\cal M}_c + {\cal M}_s,
\label{app1-1}
\end{equation}
with the condition
\begin{equation}
|E_r| \ll  |{\cal M}_c(\alpha)| \sim |{\cal M}_s(\alpha_s)|.
\label{app1-2}
\end{equation}
We then obtain
\begin{eqnarray}
\Delta E_r &=& \frac{\partial E_r}{\partial\alpha}\Delta\alpha +\frac{\partial E_r}{\partial\alpha_s}\Delta\alpha_s, \nonumber\\
&=& \frac{{\cal M}_c}{\alpha}\Delta\alpha +\frac{{\cal M}_s}{\alpha_s}\Delta\alpha_s.
\label{app1-3}
\end{eqnarray}
In deriving the second equation we assumed 
\begin{equation}
\frac{\partial {\cal M}_c}{\partial\alpha} =\frac{{\cal M}_c}{\alpha}, \quad\mbox{and}\quad \frac{\partial {\cal M}_s}{\partial\alpha_s} =\frac{{\cal M}_s}{\alpha_s},
\label{app1-4}
\end{equation}
to simplify the equations, for the moment.

According to (\ref{app1-1}) and (\ref{app1-2}) we put (\ref{app1-3}) into
\begin{equation}
\Delta E_r \approx {\cal M}_c \left( \frac{\Delta\alpha}{\alpha}-\frac{\Delta\alpha_s}{\alpha_s} \right).
\label{app1-6}
\end{equation}
Ignoring the second term yields the Coulomb-only estimate
\begin{equation}
\frac{\Delta\alpha}{\alpha} \approx \frac{\Delta E_r}{{\cal M}_c} \equiv D_{c0}.\label{app1-7}
\end{equation}
On the other hand, we notice that the right-hand side of (\ref{app1-6}) happens to vanish if 
\begin{equation}
\frac{\Delta\alpha_s}{\alpha_s} = \frac{\Delta\alpha}{\alpha},
\label{app1-5}
\end{equation}
leaving $\Delta\alpha/\alpha$ undetermined in terms of $\Delta E_r$.  (We then have to bring $E_r$ back again on the right-hand side of (\ref{app1-6}), as discussed at the end of Section 5 of [\cite{yfetal}], thus corresponding to the ``exception" mentioned toward the end of Section 3. We ignore this case at this moment.)  We may assume, however,
\begin{equation}
\frac{\Delta\alpha_s}{\alpha_s} = \xi \frac{\Delta\alpha}{\alpha},
\label{app1-8}
\end{equation}
where $\xi$ is to be determined based on the more fundamental laws of physics, as attempted in [\cite{olive},\cite{fritzsch},\cite{lang}]. By using this in (\ref{app1-6}), we obtain
\begin{equation}
\frac{\Delta\alpha}{\alpha} \approx (1-\xi)^{-1}D_{c0}.
\label{app1-9}
\end{equation}

Suppose a special relation (\ref{app1-5}) holds true within the accuracy of 10\%, for example.  This implies that (\ref{app1-8}) holds true for
\begin{equation}
\xi = 1 + \delta\xi, \quad\mbox{with}\quad |\delta\xi| \:\lsim\: 0.1.
\label{app1-11}
\end{equation}
Then (\ref{app1-9}) implies
\begin{equation}
\Biggl| \frac{1}{D_{c0}}\frac{\Delta\alpha}{\alpha}\Biggr| = |\delta\xi|^{-1} \:\gsim\: 10.
\label{app1-12}
\end{equation}
In other words, $\Delta\alpha/\alpha$ should remain close to the Coulomb-only estimate within an order of magnitude, unless the equality $\Delta\alpha/\alpha = \Delta\alpha_s/\alpha_s$ holds true to the accuracy better than 10\%.

This result may be extended to more general situations, in which ${\cal M}_c$ and ${\cal M}_s$ depend on $\alpha_s$ and $\alpha$, respectively, though then separating into ${\cal M}_c$ and ${\cal M}_s$ in (\ref{app1-1}) may not be unique.  A certain relation like (\ref{app1-5}) is expected to result in the vanishing right-hand side of an equation corresponding to (\ref{app1-6}).  It is unlikely that the relation of this kind holds true exactly in practice.\footnote{From the minimal supersymmetric standard model follows $\xi \approx 6$ at $\mu = M_Z$ [\cite{lang}], though this result is not readily extrapolated  to a much smaller $\mu$ in QCD, making it even unlikely to derive a value anywhere near unity}  Unless it does within the accuracy of 10\%, we should always expect $\Delta\alpha/\alpha$  to remain less than an order of magnitude of the Coulomb-only estimate.

\section{Distant migration of the higher resonances}
\setcounter{equation}{0}

We have so far assumed that $\Delta E_r$ is very small, much smaller than $E_r = 93.7\:{\rm meV}$ for Sm.  In fact we obtained $\Delta E_r$ as small as $10\:{\rm meV}$, thus giving $|\Delta\alpha/\alpha|\sim 10^{-8}$.  This is a right attitude as long as we try to find as small an upper bound as we can.  Now, however, the QSO result indicates a much larger value, up to 2 or 3 orders of magnitude larger.  This might raise a question if the Oklo phenomenon does in fact yield a correspondingly larger $\Delta E_r$.  One may suggest that we are looking at in the remnants of Oklo RZ a distant ``migration" of a higher resonance down nearly to the threshold of $n + ^{149}{\rm Sm}$.

This possibility was already discussed by Shlyakhter [\cite{ATOMKI}], based on a statistical argument.  Inspired by his approach, Akira Iwamoto and the present author attempt here a similar analysis on the issue, reaching a rather negative conclusion by including the observation of Gd.  We start, however, with discussing Sm first, for which the energies and the widths of the first four resonances are shown in Table \ref{table3}.

\begin{table}[htp]
\caption{First four resonances in $n+^{149}\!{\rm Sm}$. $\Gamma_n$ is the elastic width. The last line represents approximately expected time-variability of $\alpha$ obtained by the Coulomb-only estimate in units of $10^{-5}$
}
\begin{center}
\renewcommand{\arraystretch}{1.4}
\setlength\tabcolsep{5pt}
\begin{tabular}{ccccc}
Resonances & 1st & 2nd & 3rd & 4th \\
\hline
$E_r$ (meV) & 97 & 872 & 4950 & 6430 \\
$\Gamma_n$ (meV) & 0.53 & 0.74 & 2.13 & 0.72 \\
$\Gamma_{\rm tot}$ (meV) & 30 & 30 &31 & 30 \\
$\Delta\alpha/\alpha$ $(10^{-5})$ &$\sim 10^{-3}$ & 0.08 & 0.45 &0.58 \\
\hline
\end{tabular}
\end{center}
\label{table3}
\end{table}

Suppose  $\Delta E_r$ was negatively so large 2 Gys ago that one of the higher levels came down nearly at the same position as the first resonance.  We may expect a significant effect if its (shifted) energy minus $97\:{\rm meV}$ happens to be close to the threshold.  We may ignore the thermal energy $(60-90)\:{\rm meV}$ corresponding to the estimated temperature of $(200-400)^\circ{\rm C}$ in the natural reactors.  According to Table \ref{table3}, the elastic width $\Gamma_n$ is found to be proportional to the center-of-mass momentum within an order of magnitude.  We then expect the Coulomb-only estimate $\Delta\alpha/\alpha=\Delta E_r/{\cal M}_c \approx -E_r/{\cal M}_c$ to be roughly correct with a common value ${\cal M}_c \approx -1.1 \:{\rm MeV}$ for $E_r \:\ll \:{\rm MeV}$.  This is the way we have obtained the last line of Table \ref{table3}. Interesting enough, the values for the last two resonances turn out to be comparable with those reported by the QSO observations [\cite{ww2}], but with the wrong sign.

We notice, however, that $\Gamma_{\rm tot}$ is considerably smaller than the energy difference required for the shift, implying that the ``probability" of finding a shifted level that falls in the range of $\sim \Gamma_{\rm tot} \sim 30\:{\rm meV}$ around the threshold is rather small.  This conclusion seems to be corroborated by taking $^{155,157}{\rm Gd}$ into account as will be discussed.

\begin{figure}[tbh]
\vspace{-2em}
\hspace{9em}
\epsfxsize=6.cm
\epsffile{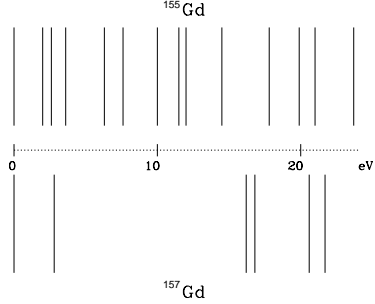}

\vspace{-3em}
\caption[]{Levels of resonances of $n+^{155}\!\!{\rm Gd}$ and $n+^{157}\!\!{\rm Gd}$, shown above and below the horizontal axis, respectively.  The scale of the energy levels is eV.  They appear to be distributed rather randomly, except for the first resonances which are nearly degenerate
}
\label{figapp2}
\end{figure}

There are many excited levels also in $n+^{155,157}\!{\rm Gd}$, some of which are illustrated in Fig. \ref{figapp2}.  The first resonance levels appear at exceptionally low energies, $26.8{\rm meV}$ and $31.4{\rm meV}$, respectively.  They are nearly degenerate.  This unique feature is shared by none of the higher resonances, though all of them show remarkably similar widths, around $100\:{\rm meV}$.
\mbox{}\\[-1em]

The analysis of the Oklo natural reactors shows, however, that considerable enhancement near the threshold appears to occur for both isotopes.  This conclusion seems to remain true even if possible significant effect of ``contamination" is included [\cite{yfetal}].   What we observe is reasonably interpreted as coming from both of the two resonances.

The inherent ambiguity coming from higher resonances as encountered for $n+^{149}\!{\rm Sm}$ is also relevant here.  This time, however, we must expect the levels to land near the thresholds ``simultaneously" in both reactions.  This requires ``squared" smallness of the probability in view of  Fig. \ref{figapp2}.  Note that the ratios of the widths to the level spacings are even smaller than those in Sm. Small probability will be ``cubic" if we combine all of the results of Sm and Gd.

All in all, it is highly unlikely that the ``discrepancy" between the Oklo constraint and the QSO result can be removed by assuming ``distant migration" of higher resonance levels of the relevant isotopes.

\section{Another 3-parameter fit with an offset}
\setcounter{equation}{0}

\begin{figure}[htb]
\vspace{-12em}
\hspace{9em}
\epsfxsize=8.cm
\epsffile{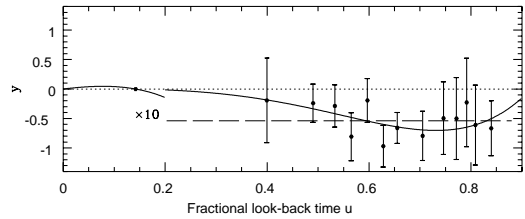}

\caption[]{The best fit of (\ref{app3-1}) for $a=0.046, b= 4.0, T= 1.307$ giving $\chi^2_{\rm red}= 1.071$ which is comparable with 1.06 for the weighted mean of [\cite{ww2}]  
}
\label{app3-fig1}
\end{figure}

We have recently found a fit with an offset parameter included,  parametrized by
\begin{equation} 
y(u) = a\left( e^{bu}\cos \left(v-v_1  \right) -\cos\left( v_1 \right)  \right),
\label{app3-1}
\end{equation} 
where $v/u= v_{\rm oklo}/u_{\rm oklo}=2\pi T^{-1}$  with $v_1$ determined by
\begin{equation}
v_1 = \tan^{-1}\left( \left( e^{-b u_{\rm oklo}} -\cos (v_{\rm oklo})   \right)/\sin (v_{\rm oklo})  \right).
\label{isoch-17}ls
\end{equation}
We easily find that $y$ defined this way vanishes both at $u=0$ and $u=u_{\rm oklo}$.  The three parameters $a, b, T$ are then determined to minimize $\chi^2$ for the QSO data.  The result is for $a=0.046, b= 4.0, T=1.307$ with the fit shown in Fig. \ref{app3-fig1}.  The resulting $\chi^2_{\rm red}=1.071$ is even smaller than 1.09 for our previous 3-parameter fit in Fig. \ref{fig7}.  

Through these analyses we also find it unlikely that the current QSO result respects another constraint at the ``meteorite time" around 4.6 Gys ago, or $u\approx 0.33$, requiring $|y|\:\lsim\: 0.025$ [\cite{olive}].  We are re-examining the basic formulation used in analyzing the decay of $^{187}{\rm Re}\rightarrow ^{187}\!\!{\rm Os}$ [\cite{olive2},\cite{yfai}].\\[1em]

\newpage

\noindent
{\large\bf References}\\

\baselineskip = .2cm
\begin{enumerate}
{\small
\item\label{naudet}See, for example, R. Naudet, {\sl Oklo: des r\'{e}acteurs necl\'{e}aires fossiles}, Collection CEA, Eyrolles, Paris, 1991

\item\label{IAEA}The Oklo Phenomenon, Proc. of a Symposium, Libreville, June, 1975 (IAEA, Vienna, 1975)

\item\label{kuroda}P.K. Kuroda, J. Chem. Phys. \textbf{25}, 781, 1295 (1956)

\item\label{nat}A.I. Shlyakhter, Nature \textbf{264}, 340 (1976)

\item\label{ATOMKI}A.I. Shlyakhter, ATOMKI Report A/1, unpublished (1983), physics/0307023

\item\label{yfetal}Y. Fujii, A. Iwamoto, T. Fukahori, T. Ohnuki, M. Nakagawa, H. Hidaka, Y. Oura and P. M\"{o}ller, Nucl. Phys. {\bf B573}, 377 (2000).  See also, hep-ph/0205206

\item\label{dd}T. Damour and F.J. Dyson, Nucl. Phys. {\bf B480}, 37 (1996)

\item\label{olive}K.A. Olive, M. Pospelov, Y.-Z. Qian, A. Coc, M. Cass\'{e} and E. Vangioni-Flam, Phys. Rev. {\bf D66}, 045022 (2002)

\item\label{ww2}M.T. Murphy, J.K. Webb, V.V. Flambaum, 
MNRAS, to be published, astro-ph/0306483

\item\label{ww3}M.T. Murphy, V.V. Flambaum et al., the present proceedings

\item\label{BM1}J.D. Barrow and J. Magueijo, Astrophys. J., {\bf 532}, L87 (2000)

\item\label{BT}J.D. Barrow and C. O'Toole, astro-ph/9904116

\item\label{wett}C. Wetterich, Phys. Lett. {\bf B561}, 10 (2003)

\item\label{bek}J.D. Bekenstein, Phys. Rev. {\bf D25}, 1527 (1982): gr-qc/0208081

\item\label{anchor}L. Anchordoqui and H. Goldberg, hep-ph/0306084

\item\label{gardner}C.L. Gardner, astro-ph/0305080

\item\label{randp}A.G. Riess {\em et al.}, Astgron. J. {\bf 116}, 1009 (1998); S. Perlmutter {\em et al}., Nature, {\bf 391}, 51 (1998)

\item\label{yfdcl}Y. Fujii, Phys. Rev. {\bf D26}, 2580 (1982)

\item\label{dolgov}D. Dolgov, Proc. Nuffield Workshop, ed. G.W. Gibbons and S.T. Siklos, Cambridge University Press, 1982

\item\label{cup}Y. Fujii and K. Maeda, {\sl The scalar-tensor theory of gravitation}, Cambridge University Press, 2003

\item\label{yfplb}Y. Fujii, Phys. Lett. {\bf B573}, 39 (2003), astro-ph/0307263 

\item\label{fritzsch}X. Calmet and H. Fritzsch, Eur. Phys. J. {\bf C24}, 639 (2002)

\item\label{lang}P. Langacker, G. Segr\`{e} and M. Strassler, Phys. Lett. {\bf B528}, 121 (2002)

\item\label{olive2}K.A. Olive, M. Pospelov, Y.-Z. Qian, G. Manh\`{e}s,
	E. Vangioni-Flam, A. Coc and  M. Cass\'{e}, astro-ph/0309252

\item\label{yfai}Y. Fujii and A. Iwamoto, Phys. Rev. Lett., to be published, hep-ph/0309087
}
\end{enumerate}

\end{document}